\documentclass[twocolumn, showpacs, prl]{revtex4-1}
\usepackage{graphicx, color}
\def\ind#1{{_{\mathrm{#1}}}}

\begin{document}

\title{Dynamics of localized modes in a composite multiferroic chain}

\author{L. Chotorlishvili$^1$, R. Khomeriki$^{2,3}$, A. Sukhov$^1$, S. Ruffo$^4$ and J. Berakdar$^1$}

\address{$^1$Institut f\"ur Physik, Martin-Luther Universit\"at Halle-Wittenberg, D-06120 Halle/Saale, Germany \\
$^2$Physics Department, Tbilisi State University, 0128 Tbilisi, Georgia \\
$^2$Max-Planck Institute for the Physics of Complex Systems,
N\"othnitzer Str. 38, 01187 Dresden, Germany \\
$^4$Dipartimento di Fisica e Astronomia and CSDC, Universit\`a di
Firenze, CNISM and INFN, via G. Sansone, 1, Sesto Fiorentino,
Italy}

\date{\today}

\begin{abstract}
In a  coupled ferroelectric/ferromagnetic system, i.e. a composite
multiferroic, the propagation of magnetic or ferroelectric
excitations across the whole structure is a key issue for
applications.  Of a special  interest is the dynamics of localized
magnetic or ferroelectric modes (LM) across the
ferroelectric-ferromagnetic interface, particularly when the LM's
carrier frequency is in the  band of the ferroelectric and in the
band gap of the ferromagnet. For a proper choice of the system's
parameters, we find that there is a threshold amplitude above
which the interface becomes transparent and a band gap
ferroelectric LM penetrates the ferromagnetic array. Below that
threshold, the LM is fully reflected. Slightly below this
transmission threshold, the addition of noise may lead to energy
transmission, provided that the noise level is not too low {nor}
too high, {{ an effect that resembles stochastic resonance}}.
These findings represent an important step towards the application
of ferroelectric and/or ferromagnetic LM-based logic.
\end{abstract}
\maketitle
\emph{Introduction}.- {Multiferroics (MF) possess coupled ferroic
(magnetic, electric, or elastic) ordering
\cite{ME-review,Single-ME,Composite-ME}. The current high interest
in MF is fueled by the impressive advances in synthesizing
composite ferroelectric (FE)/ferromagnetic (FM) nano and multi
layer structures. These  show a substantially larger multiferroic
coupling strength
\cite{ME-review,Single-ME,Composite-ME,Composite-ME, Zavaliche-11,
MeKl11} as compared to bulk matter, so-called single-phase
multiferroics \cite{ME-review,RaSp07} such as
Cr$_2$O$_3$\cite{Dz59}.}
MFs are important for addressing fundamental questions regarding
the { connection} between electronic correlation, symmetry,
magnetism, and polarization. {They also hold the promise for
qualitatively new device  concepts based on exploiting }the
magnetoelectric (ME) coupling to steer  magnetism
(ferroelectricity) via  electric (magnetic) {fields. Potential}
applications are wide and range from sensorics  and
magnetoelectric spintronics  to environmentally friendly devices
with ultra low heat dissipation \cite{ME-memory, MTJ1,MTJ2}.
Thereby, a key issue is how { efficiently} magnetic or
ferroelectric information, i.e. an initial excitation, is
transmitted in a system with a MF coupling. For instance, in a
two-phase or composite MF \cite{ME-review,RaSp07,NaBi08} such as
BaTiO$_3$/CoFe$_2$O$_4$ \cite{BoRu76},
PbZr$\ind{1-x}$Ti$\ind{x}$O$\ind{3}$/ferrites \cite{SpFi05,Fi05},
BaTiO$\ind{3}$/Fe \cite{DuJa06}, PbTiO$\ind{3}$/Fe
\cite{FeMa10,LeSa10} or BaTiO$\ind{3}$/Ni the MF coupling is
strongest at the FE/FM interface, whereas away from it the FE or
FM order is only marginally affected. { Thus, we expect} that a
ferroelectric signal triggered by an electric field in the FE part
may or may not be converted into a magnetic signal depending on
the dynamics taking place at the interface. How this transport of
information depends on the properties of the system is rarely
studied and will be addressed in this Letter. The outcome of such
a study would not only uncover the conditions for optimal signal
handling but also holds the potential for new insights into the {
multiferroic} coupling retrieved by tracing the signal dynamics.
We will focus on weakly nonlinear localized modes (LM) which are
formed by a modulation of linear excitations of the ferroelectric
and the ferromagnetic systems. {Such nonlinear modes have in
isolated FM or FE phases a series of applications in magnetic
logic, microwave signal processing, and spin electronic devices.}{
{A clear { advantage} is that LMs of a large number of elementary
excitations are very robust and have a particle-like
nature~\cite{Kos90}.} In that sense, LMs are very similar to their
topological counterparts (magnetic solitons) that have been
considered for logic operations~\cite{cowburn,apl,science}.
\begin{figure}[!b]
\centering \includegraphics[scale=.4]{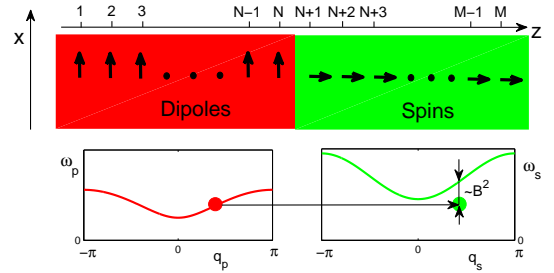}
\caption{(color online) Schematics of a chain consisting of a
ferroelectric and a ferromagnetic part coupled at the interface.
The lower panel shows a particular choice of frequency for which a
conventional localized mode is formed in the ferroelectric. In the
ferromagnetic part a bandgap localized excitation develops with a
nonlinear frequency shift proportional to $B^2$ ($B$ is the
amplitude of a magnetic band-gap localized mode). The chosen
mutual alignment of the ferroelectric polarization and the
magnetization at the interface resembles the realized
ferroelectric (BaTiO$_3$) tunnel junction with ferromagnetic (Fe)
electrodes \cite{GaBi10}.} \label{fig_0}
\end{figure}
Multiferroics offer new fascinating mechanisms for LM
dynamics~\cite{new}. For example, due to discreteness and/or
nonlinearity of the system, it may happen that the large-amplitude
excitation  frequency  falls within the gap of the linear
oscillations spectrum, as illustrated in Fig.\ref{fig_0}. Then,
the energy of the excitation would not spread over the lattice. As
well-established in studies on intrinsic LMs,
e.g.~\cite{SwBr99,SchwEn99,KoAu01,FlKl99,BaRa06,GiCh11,GiChGu11,BaRa10},
we know that, in spite of the localized energy profile, such modes
may move along the whole chain. This means that excitations
created in the ferroelectric part via an electric field can be
transmitted to the magnetic part and move there. Moreover, as it
will be shown { below,} the creation of these band-gap LMs { can}
be enhanced by { noise,} manifesting thus some analogy with
{ the} stochastic resonance phenomenon \cite{BeSu81,GaHa98}.\\
\emph{Model}.- For our { purposes,} a large ME coupling is
necessary. In this { respect,}
 new fabrication methods \cite{Fi05} for
{ the} so-called two-phase or composite multiferroics
\cite{ME-review,RaSp07,NaBi08}, as well as the realization of
ferroelectric wires, \cite{Alexe} are encouraging. Examples of
composite multiferroics BaTiO$_3$/CoFe$_2$O$_4$ \cite{BoRu76} or
PbZr$\ind{1-x}$Ti$\ind{x}$O$\ind{3}$/ferrites \cite{Fi05} are
still popular. Major research is focused on BaTiO$\ind{3}$/Fe
\cite{DuJa06}, PbTiO$\ind{3}$/Fe \cite{FeMa10,LeSa10} or
BaTiO$\ind{3}$/Ni composite multiferroics, to name but a few,
since their bulk parameters are very well known (Ref.
\cite{RaAh07} for BaTiO$\ind{3}$ and Ref. \cite{Coey10} for Fe or
Ni) as well as the misfit {of the lattices }is relatively low
\cite{FeMa10}. Relatively high ME constants \cite{FeMa10,LeSa10}
were predicted for these materials at room temperatures. A
possible
 mechanism for ME coupling at the FE/FM interface is based on screening
 effects~\cite{CaJu09}. {
We assume here { the presence of} a similar mechanism based
microscopically  on { the rearrangement} of charges and spins at
the FM/FE interface, as confirmed by other
studies~\cite{DuJa06,MeKl11}. The spin-polarized charge density
formed in the FM in the vicinity of the FM/FE
interface~\cite{CaJu09} acts with a torque on the magnetic moments
in { the FM,} resulting  in a non-collinear magnetic ordering
(similar as in \cite{nick}). Hence, electric polarization emerges
that couples the FM to the FE part. This picture yields {a} linear
ME coupling  with a pseudoscalar coupling constant .} Technically,
we describe the bulk unstrained BaTiO$\ind{3}$  by the
Ginzburg-Landau-Devonshire (GLD) potential \cite{RaAh07}. For the
discretized FE polarization ($P_n$) in a coarse-grained approach,
the  form of the GLD potential for a general phase and arbitrary
temperatures is quite involved \cite{SuJi10}. However, at room
temperature the BaTiO$\ind{3}$-crystal has an axis along which the
polarization switches (tetragonal phase). Consequently, the form
of the GLD potential reduces to the one dimensional biquadratic
potential. For the description of the magnetization ($S_k$)
dynamics in the FM, we employ the classical Heisenberg model.
$S_k$ is discretized and normalized to the saturation value of the
coarse-grained magnetization vector. With the aim { of exploring}
the feasibility of conversion of the electric excitation formed in
FE part { of the sample} into a localized spin magnetic excitation
in FM part, { we thus} employ the multiferroic model (cf. Fig.
\ref{fig_0})
\begin{eqnarray}
&H = H\ind{P}  + H\ind{S} + V\ind{SP},\\
&H\ind{P}  = \sum\limits_{n = 1}^N {\left( {\frac{\alpha_0}{2}
\left(\frac{dP_n}{dt}\right)^2 + \frac{\alpha_1}{2}P_n^2  +
\frac{\alpha_2}{4}P_n^4  + \frac{\kappa}{2}\left( {P_{n + 1}  -
P_n } \right)^2 } \right)},
\nonumber \\
&H\ind{S}  = \sum\limits_{k = N+1}^M {\left( { - J_1\vec S_k \vec
S_{k + 1}  - J_2\left( {S_k^z } \right)^2 } \right)} , \quad
V\ind{SP} = -g P_N  S_1^x, \nonumber \label{1}
\end{eqnarray}
where $H\ind{P}$ is the Hamiltonian of { the FE part of the
multiferroic system, describing $N$-interacting FE dipoles
\cite{SuJi10,GiChGu11}} { ($P_n$ and $dP_n/dt$ are conjugated
variables).} {{  $P_n$ and $\vec S_k$ stand respectively for the
deviations from the equilibrium positions of the $n$-th dipole and
the $k$-th spin vector.} At room temperature, we can choose the
polarization vector to be directed along the $x$ axis $\vec{{P}}_n
= \left({P_n,0,0}\right),\,\,\,n = 1,\ldots,N$}.
 $\alpha_0$ is a kinetic
coefficient, $\alpha\ind{1,2}$ are potential constants and
$\kappa$ is the nearest neighbor coupling constant. $H\ind{S}$
describes the { ferromagnetic chain}~\cite{Ch02}, where $J_1$ is
the nearest neighbor exchange coupling in the FM part and $J_2$ is
the uniaxial anisotropy constant. Interface effects between the
spin and the FE dipole systems are described by the dipole-spin
interaction Hamiltonian $V\ind{SP}$.

In our numerical simulations we  operate with dimensionless
quantities upon introducing  $p_n=P_n/P_{0}$, $\vec s_k=\vec
S_k/S$ and defining a dimensionless time as
$t\rightarrow\omega_0t$ ($\omega_0=\sqrt{\kappa/\alpha_0}\sim
10^{12}$ Hz). The equations governing the time evolution  of the
dipoles and the spins (except for the sites near the interface)
read
\begin{eqnarray}\label{2}
&\frac{d^{2} p_n}{d t^{2}}  =  - \alpha p_n  - \beta p_n^3 +
\left( {p_{n - 1}
-2p_n  + p_{n + 1} } \right) \\
&\frac{\partial {s}_k^\pm}{\partial t} =  \pm iJ \left[s_k^\pm
\left(s_{k - 1}^z  + s_{k + 1}^z \right)-s_k^z \left(s_{k - 1}^\pm
+ s_{k + 1}^\pm \right)\right]\pm \nonumber \\
&\pm 2iD s_k^\pm s_k^z \label{3}
\end{eqnarray}
where $n\neq N$ and $k\neq N+1$. We { have introduced} the
following dimensionless constants $\alpha=\alpha_1/\kappa$,
$\beta=\alpha_2P_{0}^2/\kappa$, $J=J_1S/\omega_0$ and
$D=J_2S/\omega_0$. For the  dipole $p_N$ and the spin $\vec s_1$
at the interface the { following equations hold}
\begin{eqnarray}\label{4}
&\frac{d^{2} p_N}{d t^{2}}  =  - \alpha p_N  - \beta p_N^3 +
\left( p_{N - 1} -2p_N  + g_s s_1^x \right), \\
&\frac{\partial {s}_1^\pm}{\partial t} =  \pm i J\left[s_1^\pm
s_{2}^z -s_1^z s_{2}^\pm\right]\pm i \left[2Ds_1^\pm s_1^z-g_p
p_Ns_1^z\right]. \nonumber
\end{eqnarray}
Here $s_k^\pm\equiv s_k^x\pm i s_k^y$, $g_s=gS/(\kappa P_{0})$ and
$g_p=gP_{0}/(S\omega_0)$.   The  evolution according to Eqs.
(\ref{2}-\ref{4})  proceeds under the constraint
$\left(s_k^x\right)^2+\left(s_k^y\right)^2+\left(s_k^z\right)^2=1$.
For the derivation of { the weakly} nonlinear envelope solutions
from eqs. (\ref{2}) and (\ref{3}) one can { rely on} the reductive
perturbation theory developed in Ref.~\cite{OiYa74,short}. One
obtains the solutions for the dipoles and the spins separately in
the following form (a detailed derivation is provided as
supplementary material to this paper):
\begin{equation}
\displaystyle p_n  = \frac{A\cos \left[ \omega_p t - q_pn +
\delta\omega_p t\right]}{\cosh\left[(n - V_p t)/\Lambda_p \right]}
\quad  s_k^\pm= \frac{B e^{ \pm i\left(\omega_s t - q_s k
+\delta\omega_s t \right)}} {\cosh \left[(k- V_s t)/\Lambda_s
 \right]} \label{5}
\end{equation}
where $A$ and $B$ are the amplitudes of the dipolar and the
magnetic { localized excitations}, respectively; $\omega_p$ and
$\omega_s$ are the frequencies of the linear excitations which
obey the following dispersion relations
\begin{equation}
\omega_p = \sqrt{\alpha  + 2\left( {1 - \cos q_p} \right)}, \quad
\omega_s=2\left[D+J(1-\cos q_s)\right], \label{6}
\end{equation}
$q_p$ and $q_s$ are the carrier wave numbers of the dipolar and
the spin excitations; $V_p=\sin q_p/\omega_p$ and $V_s=2J\sin q_s$
are the group velocities of the corresponding LMs.
The width of the dipolar and spin LMs are
\begin{equation}
\Lambda_p=\frac{1}{A}\sqrt{\frac{2\left(\omega_p^4  - \alpha ^2 -
4\alpha\right)}{3\omega_p^2\beta}}, \quad
\Lambda_s=\frac{1}{B}\sqrt{\frac{4J\cos q_s}{\omega_s}}. \label{7}
\end{equation}
The  nonlinear frequency shifts are defined as
\begin{equation}
\delta\omega_p=A^2 \frac{3\beta}{16\omega_p}, \qquad
\delta\omega_s=-B^2\frac{\omega_s } {4}. \label{8}
\end{equation}
Note that, for  wave packet transmission, the following matching condition between the frequencies
has to be fulfilled \cite{note-3}
\begin{equation}
\omega_p+\delta\omega_p=\omega_s+\delta\omega_s. \label{9}
\end{equation}
For an efficient transmission of the {LM} from { the FE into the
FM part, the widths of the {LM}} should be the same in both parts,
i.e.\ $\Lambda_p=\Lambda_s$ with the restriction $B\leq g_pA$.
If one excites { the {LM}} with a carrier frequency $\omega$ which
is located within the band of both the dipolar and the spin wave
spectrum, then the { localization} will safely penetrate from the
FE to the FM part, but some portion of the energy  will be
reflected by the interface. By changing the amplitude of { the
{LM},} one can manipulate the ratio between the transmitted and
reflected parts of the LM. In addition, the transmission is very
sensitive to the coupling constant $g$ between { the FE and the FM
parts}. We have investigated this dependence by varying only the
coupling constant $g_s$ and fixing the values of the dimensionless
parameters as follows: $\alpha=0.2, ~~ \beta=0.1, ~~ J=1,
~~D=0.6$. We  assume for simplicity $g_p=g_s$ (in general these
constants differ, depending on the { material of the samples,
but this is not an obstacle for the theory).}\\
\emph{Numerical Results}.-
{  Realistic material parameters are tabulated in full { detail}
in the supplementary material {section. There, we }provide
explicitly the relation to the normalized units which we use
below. {The essential parameters entering eq. (1) are:} the FE
potential coefficients $\alpha_1/(a^3_{\mathrm{FE}})=2.77\cdot
10^{7}$ [Vm/C], $\alpha_2/(a^3_{\mathrm{FE}})=1.7\cdot 10^8$
[Vm$^5$/C$^3$], the FE coupling coefficient
$\kappa/(a^3_{\mathrm{FE}})=1.3\cdot 10^8$ [Vm/C], the equilibrium
polarization $P_{0}=0.265$ [C/m$^2$] and the coarse-grained FE
cell size $a_{\mathrm{FE}}=1$ [nm]. { The FM exchange interaction
strength is $J_1=3.15\cdot 10^{-20}$ [J], the FM anisotropy
constant is $J_2=6.75\cdot 10^{-21}$ [J], and the ME coupling
strength is} $g\approx 10^{-21}$ [Vm$^2$].}
 Fig. \ref{fig_00} a), b), c)
show the localized energy evolution along the lattice for
different values of the coupling constant $g_p$. In graph d) the
dependence of { the} transmitted energy on the coupling constant
is { displayed, pointing out} that the transmission is maximal
when $g_p$ is in between the spin and the dipolar coupling
constants (in reduced units it is equal to 1). Spins alignments,
the topology of the excitation, and its propagation in the {chain}
at different times are displayed in the supplementary material.
\begin{figure}[!t]
\centering \includegraphics[scale=.65]{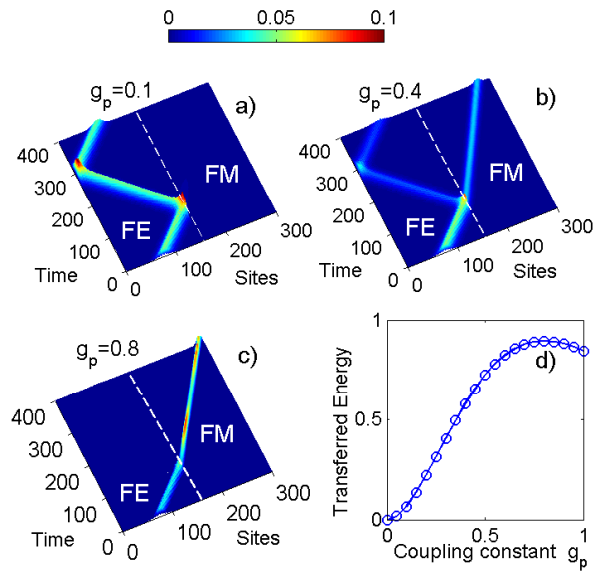}
\caption{(color online)  {Insets a), b) and c) show the time and
site dependence of the local energy. For dipoles this energy is
given by  the local values of $H\ind{P}$, and for the spins by the
local values of $g_p$ is a coupling constant indicating the
strength of the ME interface interaction.
 The graphs { point out} the LM reflection and transmission at the FE/FM interface  (white dashed line).
The  dipolar { localization} carrier wave number is chosen as
$q_p=0.4\pi$ and the dipolar { localization} amplitude is chosen
as $A=0.2$. Dipoles and spins (separated by the white { dashed
line) occupy} the sites $n=1\dots 150$ and $k=151\dots 300$,
respectively. d) { Dependence of the relative energy transferred
to the FM part (i.e. ratio of the energy in the FM part to the
total} injected energy) on the coupling constant strength $g_p$.}}
\label{fig_00}
\end{figure}
Further interesting effects arise when a band { localized
excitation} forms {with} a carrier frequency $\omega$ in the band
of the dipolar spectrum (see bottom panel of Fig. \ref{fig_0}) and
slightly below the zone boundary $\omega_s(q_s=0)$ of the spin
wave spectrum. { Then,} for small amplitudes{,} the dipolar {LM}
is totally reflected { by} the interface because it does not
resonate with any mode in the spin array. However{,} with
increasing the amplitude, there is a threshold value $A_{cr}$ (due
to the nonlinear frequency shift) above which the LM is
transmitted towards the FM part of the multiferroic chain, forming
thus a magnetic band-gap localization. Using Eqs.~(\ref{8}) and
(\ref{9}) and assuming $B=A$, one can infer the relation defining
this threshold amplitude to be
\begin{equation}
\omega_s(q_s=0)-\omega=\left(\frac{g_p^2\omega}{4}+
\frac{3\beta}{16\omega}\right)A_{cr}^2. \label{10}
\end{equation}
\begin{figure}[!t]
\centering \includegraphics[scale=.4]{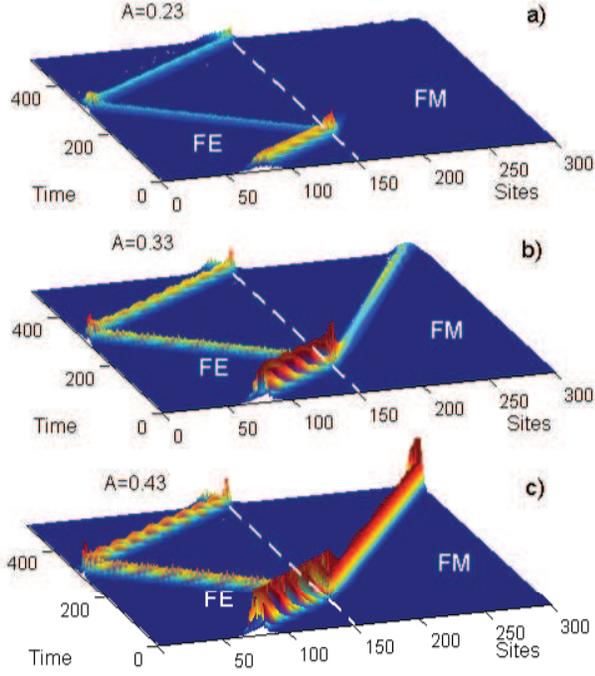}
\caption{(color online) {This figure illustrates the dependence {
on the amplitude} $A$ of the LM reflection and transmission at the
FE/FM interface. To this end we plot for different { values} of
$A$ the same quantity and choose the same parameters as in graphs
a)-c) of Fig.~\ref{fig_00} with $g_p=g_s=1$. The scale is as in
Fig. \ref{fig_00}.} } \label{fig_1}
\end{figure}
Based on this observation, we proceed with the simulations
according to Eqs. (\ref{2})-(\ref{4}) with the set of parameters
given above{. We  choose $g_p=g_s=1$ and start at $t=0$ with a
{LM} in the form of the first expression in Eq. (\ref{5}) with a
carrier wave number $q_p=0.37$. }For such a wave number the
corresponding linear frequency is $\omega=1.1856${{. This
frequency is located}} in the band gap of the spin wave spectrum
and no localization transmission { occurs} in the case of small
amplitudes{, as }it is seen from graph a) of Fig. \ref{fig_1}.
According to { relation} (\ref{10}) we can calculate the threshold
amplitude for which localization transmission emerges and find
$A_{cr}=0.22$. In the numerical results, { transmission} occurs
for the incident {LM} amplitudes $A>0.27$. This discrepancy can be
explained by the fact that localization amplitudes in different
parts of the multiferroic structure { do} not exactly coincide. In
{ panel} b) of Fig. \ref{fig_1} we display the dynamics for a
larger {LM} amplitude, i.e. $A=0.33$, and find that { localized
excitations are} formed in the ferromagnetic part as well. {{ By}
further} increasing { the} {LM} amplitude, the transmitted {
localization} takes over almost all the energy of the incident one
(graph c) of the same figure).
\begin{figure}[!t]
\centering \includegraphics[scale=.4]{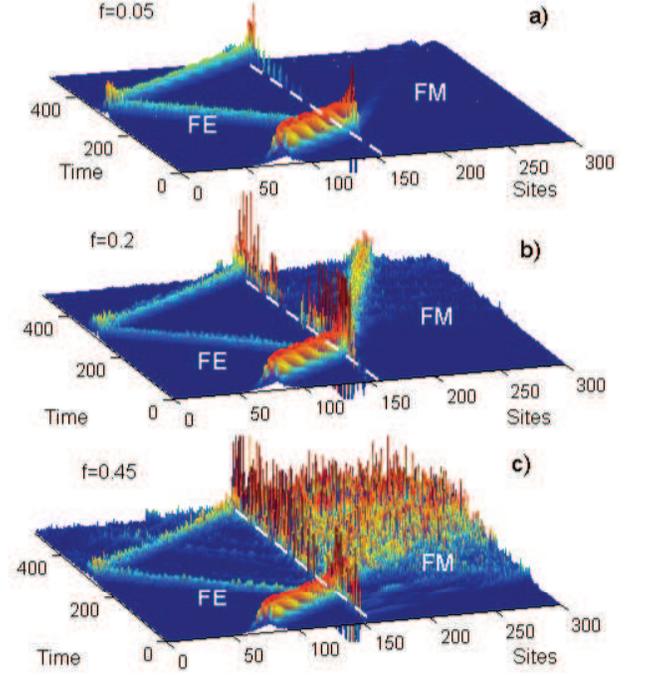}
\caption{(color online) {Influence of noise on the LM reflection
and transmission at the FE/FM interface illustrated by { realizing
similar simulations} as in Fig.~\ref{fig_1} but for $A=0.265$ and
including different noise levels as indicated on the graphs. Scale
as in Fig.\ref{fig_00}}.} \label{fig_2}
\end{figure}
If the amplitude of the incident {LM} is slightly below threshold
(here $A=0.265$),  even a small perturbation may cause { a}
transmission to { the FM} part. {Thus,} we add a term $\vec
f(t)\vec S_1$ to the Hamiltonian (\ref{1}) describing  the action
of a random magnetic field at the interface spins. $f(t)$ is
uncorrelated in time and randomly distributed in the interval
$[-f,f]$.  For small random fields, $f=0.05$, the picture is
almost the same as for zero noise (cf. upper graphs of Figs.
\ref{fig_1} and \ref{fig_2}). Increasing the noise strength to a
moderate level,  energy transmission in FM part takes place (see
graph b) of Fig. \ref{fig_2}). This stochastic resonance like
behavior is displayed in graph c)
of { Fig.} \ref{fig_2}.\\
\emph{Summary}.- As { shown} by analytical and numerical results,
in a two-phase multiferroic the magnetoelectric coupling at the
interface { determines} the conversion of an initial ferroelectric
{LM} into a ferromagnetic signal, paving thus the way for FE
and/or FM LM-based logic in multiferroics. As an essential step in
this { direction, we have} identified the conditions under which a
FE signal is converted into a FM one.

\emph{Acknowledgements}.- Consultations with Marin Alexe on the
experimental realization are gratefully  acknowledged. L.Ch., A.S.
and J.B. are  supported by DFG through SFB 762 and SU 690/1-1.
R.Kh. and S.R are funded by a joint Grant from French CNRS and
Georgian SRNSF (Grant No 09/08), R. Kh. is supported by grant
30/12 from SRNSF and S. R. acknowledge support of the contract
LORIS (ANR-10-CEXC-010-01).

\end{document}